\documentclass[preprint, aps, prd]{revtex4-1}

\def\pd{\partial}
\def\mc{\mathcal}

\usepackage[dvips]{graphicx}
\usepackage{amssymb}
\usepackage{amssymb,amsmath}
\usepackage{graphicx}
\makeatother
\begin{document}

\title{Holographic RG flows in $N=3$ Chern-Simons-Matter
theory from $N=3$ 4D gauged supergravity}

\author{Parinya Karndumri} \email[REVTeX Support:
]{parinya.ka@hotmail.com} \affiliation{String Theory and
Supergravity Group, Department of Physics, Faculty of Science,
Chulalongkorn University, 254 Phayathai Road, Pathumwan, Bangkok
10330, Thailand}

\date{\today}
\begin{abstract}
We study various supersymmetric RG flows of $N=3$
Chern-Simons-Matter theory in three dimensions by using
four-dimensional $N=3$ gauged supergravity coupled to eight vector
multiplets with $SO(3)\times SU(3)$ gauge group. The $AdS_4$ critical point preserving the full $SO(3)\times SU(3)$ provides a gravity dual of $N=3$ superconformal field theory with flavor symmetry
$SU(3)$. We study the scalar potential and identify a new
supersymmetric $AdS_4$ critical point preserving the full $N=3$
supersymmetry and unbroken $SO(3)\times U(1)$ symmetry. An analytic
RG flow solution interpolating between $SO(3)\times SU(3)$ and
$SO(3)\times U(1)$ critical points is explicitly given. We then
investigate possible RG flows from these $AdS_4$ critical points to
non-conformal field theories in the IR. All of the singularities
appearing in the IR turn out to be physically acceptable.
Furthermore, we look for supersymmetric solutions of the form
$AdS_2\times \Sigma_2$ with $\Sigma_2$ being a two-sphere or a
two-dimensional hyperbolic space and find a number of $AdS_2$
geometries preserving four supercharges with $SO(2)\times
SO(2)\times SO(2)$ and $SO(2)\times SO(2)$ symmetries.
\end{abstract}
\maketitle

\section{Introduction}
AdS$_4$/CFT$_3$ correspondence is interesting in many aspects such
as its applications in the study of M2-brane dynamics and in the
holographic dual of condensed matter physics systems. There are a
few examples of supersymmetric $AdS_4$ backgrounds with known
M-theory origins. Apart from the maximally supersymmetic $N=8$
$AdS_4\times S^7$ compactification, there is an $AdS_4$ background
with $N=3$ supersymmetry arising from a compactification of M-theory
on a tri-sasakian manifold $N^{010}$ \cite{N3_compact1}. This is a
unique solution for $2<N<8$ supersymmetry. The spectrum of the
former example has been studied in \cite{S7_spectrum} and the
massless modes can be described by the maximally $SO(8)$ gauged
supergravity constructed in \cite{N8_4D_deWit}. The lowest modes of
the latter are on the other hand encompassed in the gauged $N=3$
supergravity coupled to eight vector multiplets constructed in
\cite{N3_Ferrara}, see also \cite{N3_Ferrara2,Castellani_book}. The
holographic study of this background within the framework of $N=8$
gauged supergravity and eleven-dimensional supergravity has appeared
in many previous works, see for example
\cite{N3_AdS_CFT1,N3_AdS_CFT2,N3_AdS_CFT3}.
\\
\indent The analysis of the complete spectrum of the Kaluza-Klein
reduction of M-theory on $AdS_4\times N^{010}$ has been carried out
in \cite{N3_spectrum1}, see also \cite{N3_spectrum2}. It has been
argued that the compactification can be described by a
four-dimensional effective theory in the form of $N=3$ supergravity
coupled to eight vector multiplets with $SO(3)\times SU(3)$ gauge
group. From the AdS/CFT point of view, the $SO(3)$ and $SU(3)$
factors correspond respectively to the $SO(3)$ R-symmetry and
$SU(3)$ flavor symmetry of the dual $N=3$ superconformal field
theory (SCFT) in three dimensions with the superconformal group
$OSp(3|4)\times SU(3)$. The structure of $N=3$ multiplets and some
properties of the dual SCFT have been studied in
\cite{Ring_N3_superfield,Shadow_N3_multiplet,AdS_CFT_Tri-Sasakian,gravity_dual_CSM_theory,gravity_dual_CSM_theory1,gravity_dual_CSM_theory2}. Furthermore, a generalization to quiver gauge theories has been considered more recently in \cite{D6_AdS4_CP3,CSM_quiver1,CSM_quiver2,CSM_quiver3,CSM_quiver4,CSM_quiver5}.
\\
\indent In the present work, we are interested in exploring possible
supersymmetric solutions within four-dimensional $N=3$ gauged supergravity. The $N=3$
gauged supergravity coupled to $n$ vector multiplets has been
constructed in \cite{N3_Ferrara}. The theory contains $6n$ scalar
fields parametrizing the $SU(3,n)/SU(3)\times SU(n)\times U(1)$
coset manifold. We will focus on the case of $n=8$ which, together
with the other three vectors from the supergravity multiplet, gives
rise to eleven vector fields corresponding to a gauging of the
$SO(3)\times SU(3)$ subgroup of the global symmetry group $SU(3,8)$.
The maximally supersymmetric $AdS_4$ critical point of the resulting
gauged supergravity with all scalars vanishing is expected to describe the
$AdS_4\times N^{010}$ background of eleven-dimensional supergravity.
\\
\indent We will look for other possible
supersymmetric $AdS_4$ critical points. According to the standard
dictionary of the AdS/CFT correspondence, these should be dual to
other conformal fixed points in the IR of the UV $N=3$ SCFT with the $SU(3)$ flavor symmetry. We find that indeed there exists a
non-trivial supersymmetric $AdS_4$ critical point with $SO(3)\times
U(1)$ symmetry and unbroken $N=3$ supersymmetry. We will also
investigate holographic RG flows from the UV $N=3$ SCFT to
non-conformal field theories by looking for domain wall solutions
interpolating between the $AdS_4$ critical points and some singular
domain wall geometries in the IR.
\\
\indent Finally, we will look for supersymmetric $AdS_2\times
\Sigma_2$ solutions with $\Sigma_2$ being a Riemann surface. Like
the higher-dimensional solutions, these solutions should be
interpreted as twisted compactifications of the $N=3$ SCFTs in three
dimensions to one dimensional space-time. These results could be
interesting both in the holography of three-dimensional SCFTs and in
the context of AdS$_2$/CFT$_1$ correspondence which plays an
important role in black hole physics, see for example
\cite{AdS2_BH_Sen} and \cite{AdS2_CFT1_Sen}. Along this line, the topologically twisted indices for these theories on $S^2$ have been computed in \cite{AdS2_BH1,AdS2_BH2}. These results can be used to find the microscopic entropy of $AdS_4$ black holes by following the approach of \cite{AdS2_BH3}.
\\
\indent The paper is organized as follow. In section \ref{N3theory},
we review $N=3$ gauged supergravity in four dimensions coupled to
eight vector multiplets. In section \ref{AdS4_flow}, we will give an
explicit parametrization of $SU(3,8)/SU(3)\times SU(8)\times U(1)$
coset and study the scalar potential for the $SO(3)_{\textrm{diag}}$
singlet scalars and identify possible supersymmetric vacua. An
analytic RG flow from the UV $SO(3)\times SU(3)$ SCFT to a new IR
fixed point with residual symmetry $SO(3)_{\textrm{diag}}\times
U(1)$ is also given. We then move to possible supersymmetric RG
flows to non-conformal field theories in section
\ref{Non-conformal}. Supersymetric $AdS_2$ backgrounds obtained from
twisted compactifications of $AdS_4$ on a Riemann surface are given
in section \ref{Lower-dimensional flow}. Some conclusions and
comments on the results reported in this paper are presented in
section \ref{conclusions}.
\section{$N=3$ gauged supergravity coupled to vector multiplets}\label{N3theory}
In order to fix the notation and describe the relevant framework
from which all the results are obtained, we will give a brief
description of $N=3$ gauged supergravity coupled to $n$ vector
multiplets and finally restrict ourselves to the case of $n=8$. The
theory has been constructed in \cite{N3_Ferrara} by using the
geometric group manifold approach. For the present work, the
space-time bosonic Lagrangian and supersymmetry transformations of
fermionic component fields are sufficient. Therefore, we will focus
only on these parts. The interested reader can find a more detailed
construction in \cite{N3_Ferrara}.
\\
\indent In four dimensions, the matter fields allowed in $N=3$
supersymmetry are given by the fields in a vector multiplet with the
following field content
\begin{displaymath}
(A_\mu, \lambda_A, \lambda, z_A).
\end{displaymath}
Indices $A,B,\ldots=1,2,3$ denote the fundamental representation of
the $SU(3)_R$ part of the full $SU(3)_R\times U(1)_R$ R-symmetry.
Each vector multiplet contains a vector field $A_\mu$, four spinor
fields $\lambda$ and $\lambda_A$ which are respectively singlet and
triplet of $SU(3)_R$, and three complex scalars $z_A$ in the
fundamental of $SU(3)_R$. For $n$ vector multiplets, we use indices
$i,j,\ldots =1,\ldots, n$ to label each of them. Space-time and
tangent space indices will be denoted by $\mu,\nu,\ldots$ and
$a,b,\ldots$, respectively. In contrast to the construction in
\cite{N3_Ferrara}, we will use the metric signature $(-+++)$
throughout this paper.
\\
\indent The $N=3$ supergravity multiplet consists of the following
fields
\begin{displaymath}
(e^a_\mu, \psi_{\mu A}, A_{\mu A}, \chi).
\end{displaymath}
$e^a_\mu$ is the usual graviton, and $\psi_{\mu A}$ are three
gravitini. The gravity multiplet also contains three vector fields
$A_{\mu A}$ and an $SU(3)_R$ singlet spinor field $\chi$. It should
be noted that the fermions are subject to the chirality projection
conditions
\begin{equation}
\psi_{\mu A}=\gamma_5\psi_{\mu A},\qquad \chi=\gamma_5\chi,\qquad
\lambda_A=\gamma_5\lambda_A,\qquad \lambda=-\gamma_5\lambda\, .
\end{equation}
These also imply $\psi_\mu^A=-\gamma_5\psi_\mu^A$ and
$\lambda^A=-\gamma_5\lambda^A$.
\\
\indent With $n$ vector multiplets, there are $3n$ complex scalar
fields $z_A^{\phantom{A}i}$ living in the coset space
$SU(3,n)/SU(3)\times SU(n)\times U(1)$. These scalars are
conveniently parametrized by the coset representative
$L(z)_\Lambda^{\phantom{\Lambda}\Sigma}$. From now on, indices
$\Lambda, \Sigma, \ldots$ will take the values $1,\ldots, n+3$. The
coset representative transforms under the global $G=SU(3,n)$ and the
local $H=SU(3)\times SU(n)\times U(1)$ symmetries by a left and
right multiplications, respectively. It is convenient to split the
index corresponding to $H$ transformation as $\Sigma=(A,i)$, so we
can write
$L_\Lambda^{\phantom{\Lambda}\Sigma}=(L_\Lambda^{\phantom{\Lambda}A},L_\Lambda^{\phantom{\Lambda}i})$.
\\
\indent Together with three vector fields from the gravity
multiplet, there are $n+3$ vectors which, accompanying by their
magnetic dual, transform as the fundamental representation
$\mathbf{n+3}$ of the global symmetry $SU(3,n)$. These vector fields
will be grouped together by a single notation $A_\Lambda=(A_A,A_i)$.
From the result of \cite{N3_Ferrara}, after gauging, a particular
subgroup of $SO(3,n)\subset SU(3,n)$ becomes a local symmetry. The
corresponding non-abelian gauge field strengths are given by
\begin{equation}
F_\Lambda=dA_\Lambda+f_{\Lambda}^{\phantom{\Lambda}\Sigma\Gamma}A_\Sigma\wedge
A_{\Gamma}
\end{equation}
where $f_{\Lambda\Sigma}^{\phantom{\Lambda\Sigma}\Gamma}$ denote the
structure constants of the gauge group. The gauge generators
$T_\Lambda$ satisfy
\begin{equation}
\left[T_\Lambda,T_\Sigma\right]=f_{\Lambda\Sigma}^{\phantom{\Lambda\Sigma}\Gamma}T_\Gamma\,
.
\end{equation}
Indices on $f_{\Lambda\Sigma}^{\phantom{\Lambda\Sigma}\Gamma}$ are
raised and lowered by the $SU(3,n)$ invariant tensor
\begin{equation}
J_{\Lambda\Sigma}=J^{\Lambda\Sigma}=(\delta_{AB},-\delta_{ij})
\end{equation}
which will become the Killing form of the gauge group in the
presence of gauging.
\\
\indent As pointed out in \cite{N3_Ferrara}, one of the possible
gauge groups takes the form of $SO(3)\times H_n$ with $SO(3)\subset
SU(3)$ and $H_n$ being an $n$-dimensional subgroup of $SO(n)\subset
SU(n)$. In this case, only electric vector fields participate in the
gauging. As a general requirement, gaugings consistent with
supersymmetry impose the condition that $f_{\Lambda\Sigma\Gamma}$
obtained from the gauge structure constants via
$f_{\Lambda\Sigma\Gamma}=f_{\Lambda\Sigma}^{\phantom{\Lambda\Sigma}\Gamma'}J_{\Gamma'\Gamma}$
are totally antisymmetric. In the present paper, we are interested
only in this compact gauge group with a particular choice of
$H_8=SU(3)$ with
$f_{\Lambda\Sigma}^{\phantom{\Lambda\Sigma}\Gamma}=(g_1\epsilon_{ABC},g_2
f_{ijk})$. This choice clearly satisfies the consistency condition.
$f_{ijk}$ denote the $SU(3)$ structure constants while $g_1$ and
$g_2$ are $SO(3)\times SU(3)$ gauge couplings. The independent,
non-vanishing, components of $f_{ijk}$ can be explicitly written as
\begin{eqnarray}
f_{123}&=&1,\qquad f_{147}=f_{246}=f_{257}=f_{345}=\frac{1}{2},\nonumber \\
f_{156}&=&f_{367}=-\frac{1}{2},\qquad
f_{458}=f_{678}=\frac{\sqrt{3}}{2}\, .
\end{eqnarray}
Other possible gauge groups will be explored in the forthcoming
paper \cite{N3_4D_gauging}.
\\
\indent The bosonic Lagrangian of the resulting gauged supergravity
can be written as
\begin{eqnarray}
e^{-1}\mc{L}&=&\frac{1}{4}R-\frac{1}{2}P_\mu^{iA}P^\mu_{Ai}-a^{\Lambda\Sigma}F^+_{\Lambda\mu\nu}F^{+\mu\nu}_\Sigma
-\bar{a}^{\Lambda\Sigma}F^-_{\Lambda\mu\nu}F^{-\mu\nu}_\Sigma\nonumber \\
&
&-\frac{i}{2}e^{-1}\epsilon^{\mu\nu\rho\sigma}(a^{\Lambda\Sigma}F^+_{\Lambda\mu\nu}-\bar{a}^{\Lambda\Sigma}
F^-_{\Lambda\mu\nu})F_{\Sigma\rho\sigma}-V\, .
\end{eqnarray}
We have translated the first order Lagrangian in the differential
form language given in \cite{N3_Ferrara} to the usual space-time
Lagrangian. In addition, we have multiplied the whole Lagrangian by
a factor of $3$. This results in a factor of $3$ in the scalar
potential compared to that given in \cite{N3_Ferrara}.
\\
\indent Before giving the definitions of all quantities appearing in
the above Lagrangian, we will present the fermionic supersymmetry
transformations read off from the rheonomic parametrization of the
fermionic curvatures as follow
\begin{eqnarray}
\delta \psi_{\mu A}&=&D_\mu \epsilon_A-2\epsilon_{ABC}G^B_{\mu\nu}\gamma^\nu\epsilon^C+S_{AB}\gamma_\mu \epsilon^B,\\
\delta \chi &=&-\frac{1}{2}G^A_{\mu\nu}\gamma^{\mu\nu}\epsilon_A+\mc{U}^A\epsilon_A,\\
\delta \lambda_i&=&-P_{i\mu}^{\phantom{i}A}\gamma^\mu \epsilon_A+\mc{N}_{iA}\epsilon^A,\\
\delta
\lambda_{iA}&=&-P_{i\mu}^{\phantom{i}B}\gamma^\mu\epsilon_{ABC}\epsilon^C-G_{i\mu\nu}\gamma^{\mu\nu}\epsilon_A
+\mc{M}_{iA}^{\phantom{iA}B}\epsilon_B\, .
\end{eqnarray}
\indent From the coset representative, we can define the
Mourer-Cartan one-form
\begin{equation}
\Omega_{\Lambda}^{\phantom{\Lambda}\Pi}=(L^{-1})_\Lambda^{\phantom{\Lambda}\Sigma}dL_\Sigma^{\phantom{\Sigma}\Pi}
+(L^{-1})_\Lambda^{\phantom{\Lambda}\Sigma}f_{\Sigma}^{\phantom{\Sigma}\Omega\Gamma}A_\Omega
L_\Gamma^{\phantom{\Gamma}\Pi}\, .
\end{equation}
The inverse of $L_\Lambda^{\phantom{\Lambda}\Sigma}$ is related to
the coset representative via the following relation
\begin{equation}
(L^{-1})_\Lambda^{\phantom{\Lambda}\Sigma}=J_{\Lambda\Pi}J^{\Sigma\Delta}(L_\Delta^{\phantom{\Delta}\Pi})^*\,
.
\end{equation}
The component $\Omega_i^{\phantom{i}A}=(\Omega_A^{\phantom{A}i})^*$
gives the vielbein $P_i^{\phantom{i}A}$ of the $SU(3,n)/SU(3)\times
SU(n)\times U(1)$ coset. Other components give the composite
connections $(Q_A^{\phantom{A}B},Q_i^{\phantom{i}j},Q)$ for
$SU(3)\times SU(n)\times U(1)$ symmetry
\begin{equation}
\Omega_A^{\phantom{A}B}=Q_A^{\phantom{A}B}-n\delta^B_AQ,\qquad
\Omega_i^{\phantom{i}j}= Q_i^{\phantom{i}j}+3\delta^j_iQ\, .
\end{equation}
It should be noted that $Q_A^{\phantom{A}A}=Q_i^{\phantom{i}i}=0$.
\\
\indent The covariant derivative for $\epsilon_A$ is defined by
\begin{equation}
D
\epsilon_A=d\epsilon_A+\frac{1}{4}\omega^{ab}\gamma_{ab}\epsilon_A+Q_A^{\phantom{A}B}\epsilon_B+\frac{1}{2}nQ\,
.
\end{equation}
The scalar matrices $S_{AB}$, $\mc{U}^A$, $\mc{N}_{iA}$ and
$\mc{M}_{iA}^{\phantom{iA}B}$ are given in terms of the ``boosted
structure constants'' $C^\Lambda_{\phantom{\Lambda}\Pi\Gamma}$ as
follow
\begin{eqnarray}
S_{AB}&=&\frac{1}{4}\left(\epsilon_{BPQ}C_A^{\phantom{A}PQ}+\epsilon_{ABC}C_M^{\phantom{M}MC}\right)\nonumber \\
&=&\frac{1}{8}\left(C_A^{\phantom{A}PQ}\epsilon_{BPQ}+C_B^{\phantom{A}PQ}\epsilon_{APQ}\right),\nonumber \\
\mc{U}^A&=&-\frac{1}{4}C_M^{\phantom{A}MA},\qquad \mc{N}_{iA}=-\frac{1}{2}\epsilon_{APQ}C_i^{\phantom{A}PQ},\nonumber \\
\mc{M}_{iA}^{\phantom{iA}B}&=&\frac{1}{2}(\delta_A^BC_{iM}^{\phantom{iM}M}-2C_{iA}^{\phantom{iA}B})
\end{eqnarray}
where
\begin{equation}
C^\Lambda_{\phantom{\Lambda}\Pi\Gamma}=L_{\Lambda'}^{\phantom{\Lambda}\Lambda}
(L^{-1})_{\Pi}^{\phantom{\Lambda}\Pi'}(L^{-1})_{\Gamma}^{\phantom{\Lambda}\Gamma'}
f_{\Pi'\Gamma'}^{\phantom{\Pi'\Gamma'}\Lambda'}\qquad
\textrm{and}\qquad
C_\Lambda^{\phantom{\Lambda}\Pi\Gamma}=J_{\Lambda\Lambda'}J^{\Pi\Pi'}J^{\Gamma\Gamma'}
(C^{\Lambda'}_{\phantom{\Lambda}\Pi'\Gamma'})^*
\end{equation}
\indent With all these definitions, the scalar potential can be
written as
\begin{eqnarray}
V&=&-2S_{AC}S^{CM}+\frac{2}{3}\mc{U}_A\mc{U}^A+\frac{1}{6}\mc{N}_{iA}\mc{N}^{iA}
+\frac{1}{6}\mc{M}^{iB}_{\phantom{iB}A}\mc{M}_{iB}^{\phantom{iB}A}\nonumber \\
&=&\frac{1}{8}|C_{iA}^{\phantom{iA}B}|^2+\frac{1}{8}|C_i^{\phantom{A}PQ}|^2-\frac{1}{4}
\left(|C_A^{\phantom{A}PQ}|^2-|C_P|^2\right)
\end{eqnarray}
with $C_P=-C_{PM}^{\phantom{PM}M}$.
\\
\indent We now come to the gauge fields. The self-dual and
antiself-dual field strengths are defined by
\begin{equation}
F^\pm_{\Lambda ab}=F_{\Lambda ab}\mp\frac{i}{2}\epsilon_{abcd}
F^{cd}_\Lambda
\end{equation}
with $\frac{1}{2}\epsilon_{abcd}F^{\pm cd}_{\Lambda}=\pm i
F_{\Lambda ab}^\pm$ and $F_{\Lambda ab}^\pm=(F_{\Lambda ab}^\mp)^*$.
The explicit form of the symmetric matrix $a_{\Lambda\Sigma}$ in
term of the coset representative is quite involved. We will not
repeat it here, but the interested reader can find a detailed
discussion in the appendix of \cite{N3_Ferrara}.
\\
\indent Finally, the field strengths appearing in the supersymmetry
transformations are given in terms of $F^{\pm}_{\Lambda \mu\nu}$ by
\begin{equation}
G^i_{\mu\nu}=-\frac{1}{2}M^{ij}(L^{-1})_j^{\phantom{j}\Lambda}F^-_{\Lambda\mu\nu},\qquad
G^A_{\mu\nu}=\frac{1}{2}M^{AB}(L^{-1})_B^{\phantom{B}\Lambda}F^+_{\Lambda\mu\nu}
\end{equation}
where $M^{ij}$ and $M^{AB}$ are respectively inverse matrices of
\begin{equation}
M_{ij}=(L^{-1})_i^{\phantom{j}\Lambda}(L^{-1})_j^{\phantom{j}\Pi}J_{\Lambda\Pi}\qquad
\textrm{and}\qquad
M_{AB}=(L^{-1})_A^{\phantom{A}\Lambda}(L^{-1})_B^{\phantom{B}\Pi}J_{\Lambda\Pi}
\end{equation}
\indent In subsequent sections, we will study supersymmetric
solutions to this gauged supergravity with $SO(3)\times SU(3)$ gauge
group.

\section{Flows to $SO(3)_{\textrm{diag}}\times U(1)$ IR fixed point with $N=3$ supersymmetry}\label{AdS4_flow}
We now consider the case of $n=8$ vector multiplets and $SO(3)\times
SU(3)$ gauge group. There are $48$ scalars transforming in
$(\mathbf{3},\bar{\mathbf{8}})+(\bar{\mathbf{3}},\mathbf{8})$
representation of the local symmetry $SU(3)\times SU(8)$. It is
efficient and more convenient to study the scalar potential on a
particular submanifold of the full $SU(3,8)/SU(3)\times SU(8)\times
U(1)$ coset space. This submanifold consists of all scalars which
are singlets under a particular subgroup of the full gauge group
$SO(3)\times SU(3)$. All vacua found on this submanifold are
guaranteed to be vacua on the full scalar manifold by a simple group
theory argument \cite{warner}.

\subsection{Supersymmetric $AdS_4$ critical points}
In term of the dual $N=3$ SCFT, the $SO(3)$ part of the full gauge
group corresponds to the R-symmetry of $N=3$ supersymmetry in three
dimensions while the $SU(3)$ part plays the role of the global
symmetry. There are no singlet scalars under the $SO(3)$ R-symmetry.
In order to have $SO(3)$ symmetry, we then consider scalars
invariant under a diagonal $SO(3)$ subgroup of $SO(3)\times
SO(3)\subset SO(3)\times SU(3)$.
\\
\indent Before going to the detail of an explicit parametrization,
we first introduce an element of $11\times 11$ matrices
\begin{equation}
(e_{\Lambda\Sigma})_{\Pi\Gamma}=\delta_{\Lambda\Pi}\delta_{\Sigma\Gamma}\,
.
\end{equation}
The $SO(3)\times SU(3)$ gauge generators can be obtained from the
structure constant
$(T_\Lambda)_{\Pi}^{\phantom{\Pi}\Gamma}=f_{\Lambda\Pi}^{\phantom{\Lambda\Pi}\Gamma}$.
Accordingly, the $SO(3)$ part is generated by
$(T^{(1)}_A)_{\Pi}^{\phantom{\Pi}\Gamma}=f_{A\Pi}^{\phantom{\Lambda\Pi}\Gamma}$,
$A=1,2,3$, and the $SU(3)$ generators are given by
$(T^{(2)}_i)_{\Pi}^{\phantom{\Pi}\Gamma}=f_{i+3,\Pi}^{\phantom{i+3,\Pi}\Gamma}$,
$i=1,\ldots, 8$. The $SO(3)_{\textrm{diag}}$ is then generated by
$(T^{(1)}_A)_{\Pi}^{\phantom{\Pi}\Gamma}+(T^{(2)}_A)_{\Pi}^{\phantom{\Pi}\Gamma}$.
\\
\indent Under $SU(3)\rightarrow SO(3)\times U(1)$, we have the
branching
\begin{equation}
\mathbf{8}=\mathbf{3}_0+\mathbf{1}_0+\mathbf{2}_3+\mathbf{2}_{-3}\,
.\label{SU3_8_decom}
\end{equation}
This implies that the $48$ scalars transform under
$SO(3)_{\textrm{diag}}\times U(1)$ as
\begin{equation}
2\times
\left[\mathbf{3}_0\times(\mathbf{3}_0+\mathbf{1}_0+\mathbf{2}_{3}+\mathbf{2}_{-3})\right]
=2\times
(\mathbf{1}_0+\mathbf{3}_0+\mathbf{5}_0+\mathbf{2}_3+\mathbf{4}_3+\mathbf{2}_{-3}+\mathbf{4}_{-3}).\label{scalar_decom_SO3}
\end{equation}
A factor of $2$ comes from the fact that both
$(\mathbf{3},\bar{\mathbf{8}})$ and $(\bar{\mathbf{3}},\mathbf{8})$
of $SU(3)\times SU(8)$ become $(\mathbf{3},\mathbf{8})$ under
$SO(3)\times SU(3)$. We see that there are two
$SO(3)_{\textrm{diag}}$ singlets. These correspond to the $SU(3,8)$
non-compact generators
\begin{eqnarray}
\hat{Y}_1&=&e_{14}+e_{41}+e_{25}+e_{52}+e_{36}+e_{63},\nonumber \\
\hat{Y}_2&=&-ie_{14}+ie_{41}-ie_{25}+ie_{52}-ie_{36}+ie_{63}\, .
\end{eqnarray}
These two generators are non-compact generators of
$SL(2,\mathbb{R})\subset SU(3,8)$ commuting with
$SO(3)_{\textrm{diag}}$. The $SO(2)$ compact generator of this
$SL(2,\mathbb{R})$ is given by
\begin{equation}
J=\textrm{diag}(2i\delta^{AB},-2i\delta^{i+3,j+3},0,0,0,0,0),\qquad
i,j=1,2,3\, .
\end{equation}
From \eqref{scalar_decom_SO3}, it should be noted that the two
singlets are uncharged under the $U(1)$ factor from $SU(3)$.
Therefore, the full symmetry of $\hat{Y}_{1,2}$ is in fact
$SO(3)_{\textrm{diag}}\times U(1)$.
\\
\indent By using an Euler angle parametrization of
$SL(2,\mathbb{R})/SO(2)\sim SO(2,1)/SO(2)\sim SU(1,1)/U(1)$, we
parametrize the coset representative by
\begin{equation}
L=e^{\varphi J}e^{\lambda \hat{Y}_1}e^{-\varphi J}\, .
\end{equation}
The resulting scalar potential can be written as
\begin{eqnarray}
V&=&-\frac{3}{64}e^{-6\lambda}\left[(1+e^{4\lambda})\left[(1+e^{2\lambda})^4g_1^2+(e^{2\lambda}-1)^4g_2^2\right]\right.\nonumber \\
& &\left.+2(e^{4\lambda}-1)^3\cos (4\varphi)g_1g_2\right].
\end{eqnarray}
The above potential admits two supersymmetric $AdS_4$ critical
points. The first one is a trivial critical point, preserving the
full $SO(3)\times SU(3)$ symmetry, with all scalars vanishing
\begin{equation}
\lambda=\varphi=0,\qquad V_0=-\frac{3}{2}g_1^2
\end{equation}
where $V_0$ is the value of the potential at the critical point, the
cosmological constant. This $AdS_4$ critical point should be identified with a compactification of M-theory on $N^{010}$ manifold and dual to an
$N=3$ SCFT in three dimensions with $SU(3)$ flavor symmetry. In the
present convention, the $AdS_4$ radius $L$ is related to the value
of the cosmological constant by
\begin{equation}
L^2=-\frac{3}{2V_0}=\frac{1}{g_1^2}\, .
\end{equation}
At this critical point, all of the $48$ scalars have $m^2L^2=-2$
in agreement with the spectrum of M-theory on $AdS_4\times N^{010}$.
These scalars are dual to operators of dimension $\Delta=1,2$ in the
dual SCFT.
\\
\indent Another supersymmetric critical point is given by
\begin{equation}
\varphi=0,\qquad \lambda=\frac{1}{2}\ln
\left[\frac{g_2-g_1}{g_2+g_1}\right],\qquad
V_0=-\frac{3g_1^2g_2^2}{2(g_2^2-g_1^2)}\, .
\end{equation}
This critical point is an $AdS_4$ critical point for $g_2^2>g_1^2$
as required by the reality of $\lambda$. That this critical point
preserves supersymmetry can be checked from the supersymmetry
transformations given in the next subsection. The $AdS_4$ radius can
be found to be
\begin{equation}
L^2=\frac{g_2^2-g_1^2}{g_1^2g_2^2}\, .
\end{equation}
More precisely, there are many critical points, equivalent to the
one given above, with $\sin(4\varphi_0)=0$ or
$\varphi=\frac{n\pi}{4}$, $n\in \mathbb{Z}$. At this critical point,
we can determine the full scalar masses as shown in table
\ref{table1}.
\begin{table}[h]
\centering
\begin{tabular}{|c|c|c|}
  \hline
  $SO(3)_{\textrm{diag}}\times U(1)$ representations & $m^2L^2\phantom{\frac{1}{2}}$ & $\Delta$  \\ \hline
  $\mathbf{1}_0$ & $4$, $-2$ & $4$, $(1,2)$  \\
  $\mathbf{2}_3$ & $0_{(\times 2)}$, $-2_{(\times 2)}$  & $3$, $(1,2)$ \\
  $\mathbf{2}_{-3}$ & $0_{(\times 2)}$, $-2_{(\times 2)}$  & $3$, $(1,2)$ \\
  $\mathbf{3}_0$ & $0_{(\times 3)}$, $-2_{(\times 3)}$  & $3$, $(1,2)$ \\
  $\mathbf{4}_3$ & $-\frac{9}{4}_{(\times 4)}$, $-2_{(\times 4)}$  & $\frac{3}{2}$, $(1,2)$ \\
  $\mathbf{4}_{-3}$ & $-\frac{9}{4}_{(\times 4)}$, $-2_{(\times 4)}$  & $\frac{3}{2}$, $(1,2)$ \\
  $\mathbf{5}_0$ & $-2_{(\times 10)}$  & $(1,2)$ \\
  \hline
\end{tabular}
\caption{Scalar masses at the $N=3$ supersymmetric $AdS_4$ critical
point with $SO(3)_{\textrm{diag}}\times U(1)$ symmetry and the
corresponding dimensions of the dual operators}\label{table1}
\end{table}
\\
\indent From the table, we see seven massless scalars corresponding
to Goldstone bosons of the symmetry breaking of $SO(3)\times SU(3)$
to $SO(3)_{\textrm{diag}}\times U(1)$. The singlet scalar $\lambda$
is dual to an irrelevant operator of dimension $4$ at this critical
point while $\varphi$ is still dual to a relevant operator of
dimension $\Delta=1,2$. It should also be noted that all the masses
satisfy the BF bound as expected for a supersymmetric critical
point.
\\
\indent There is also a non-supersymmetric critical point, but we
will not give its location and value of the cosmological constant
here due to its complexity.

\subsection{A supersymmetric RG flow}
In this subsection, we will find a supersymmetric domain wall
solution interpolating between two $AdS_4$ critical points
identified previously. In order to do this, we will set up the
corresponding BPS equations by setting the supersymmetry
transformations of fermions to zero. The non-vanishing bosonic
fields are the metric and $SO(3)_{\textrm{diag}}$ singlet scalars.
\\
\indent We adopt the standard domain wall ansatz for the
four-dimensional metric
\begin{equation}
ds^2=e^{2A(r)}dx_{1,2}^2+dr^2
\end{equation}
with $dx^2_{1,2}$ being the flat Minkowski metric in three
dimensions. We will use the same convention as in
\cite{warner_Janus}. All spinors will be written as chiral projected
Majorana spinors. For example, we have
\begin{equation}
\epsilon_A=\frac{1}{2}(1+\gamma_5)\tilde{\epsilon}^A,\qquad
\epsilon^A=\frac{1}{2}(1-\gamma_5)\tilde{\epsilon}^A
\end{equation}
where $\tilde{\epsilon}^A$ is a Majorana spinor. In this Majorana
representation, all of the gamma matrices $\gamma^a$ are real while
$\gamma_5=i\gamma_0\gamma_1\gamma_2\gamma_3$ is purely imaginary. As
a consequence, $\epsilon^A$ and $\epsilon_A$ are simply related by a
complex conjugation, $\epsilon_A=(\epsilon^A)^*$.
\\
\indent In the present case, it turns out that
$C_{M}^{\phantom{M}MA}=0$. Therefore, the variation $\delta \chi$ is
identically zero. To satisfy the conditions $\delta\lambda_i=0$ and
$\delta\lambda_{iA}=0$, we impose the following projector
\begin{equation}
\gamma^{\hat{r}}\epsilon_A=e^{i\Lambda}\epsilon^A
\end{equation}
which implies $\gamma^{\hat{r}}\epsilon^A=e^{-i\Lambda}\epsilon_A$.
With this projector, the conditions $\delta \psi_{\mu A}=0$, for
$\mu=0,1,2$, reduce to a single condition
\begin{equation}
A'e^{i\Lambda}-\mc{W}=0\label{A_eq1}
\end{equation}
where $'$ is used to denote the $r$-derivative. The
``superpotential'' $\mc{W}$ is related to the eigenvalues of
$S_{AB}$. It turns out that in the present case, $S_{AB}$ is
diagonal
\begin{equation}
S_{AB}=\frac{1}{2}\mc{W}\delta_{AB}\, .
\end{equation}
This would imply unbroken $N=3$ supersymmetry provided that the
conditions $\delta\lambda_i=0$ and $\delta\lambda_{iA}=0$ can be
satisfied. The explicit form of $\mc{W}$ is given by
\begin{eqnarray}
\mc{W}&=&-\frac{1}{8}e^{-3\lambda}\left[\left[(1+e^{2\lambda})^3g_1+(e^{2\lambda}-1)^3g_2\right]\cos(2\varphi)
\right.\nonumber
\\
&
&\left.+i\left[(1+e^{2\lambda})^3g_1-(e^{2\lambda}-1)^3g_2\right]\sin(2\varphi)\right].
\end{eqnarray}
\indent By writing $\mc{W}=|\mc{W}|e^{i\omega}\equiv We^{i\omega}$,
the imaginary part of equation \eqref{A_eq1} gives rise to the
relation
\begin{equation}
e^{i\Lambda}=\pm e^{i\omega}\, .
\end{equation}
On the other hand, $\delta \lambda_i=0$ and $\delta \lambda_{iA}=0$
equations reduce to two independent equations that can be written as
\begin{equation}
\lambda'-\frac{1}{3}e^{-i\Lambda}\frac{\pd \mc{W}}{\pd \lambda}\pm
ie^{-2\lambda}(e^{4\lambda}-1)\varphi'=0\, .
\end{equation}
These two equations imply $\varphi'=0$ or $\varphi=\varphi_0$ with
$\varphi_0$ being a constant. It turns out that consistency with the
field equations require $\sin(4\varphi_0)=0$ or
$\varphi_0=\frac{n\pi}{4}$, $n\in \mathbb{Z}$. To make the solution
interpolates between the two critical points, we will set
$\varphi_0=0$.
\\
\indent With this choice, $\mc{W}$ is real, and the phase factor
$e^{i\Lambda}$ is simply given by
\begin{equation}
e^{i\Lambda}=\pm 1\, .
\end{equation}
We can finally write down all the relevant BPS equations as
\begin{eqnarray}
\lambda'&=&\mp \frac{1}{8}e^{-3\lambda}(e^{4\lambda}-1)\left[(1+e^{2\lambda})g_1+(e^{2\lambda}-1)g_2\right],\label{lambda_eq}\\
A'&=&\pm
\frac{1}{8}e^{-3\lambda}\left[(1+e^{2\lambda})^3g_1+(e^{2\lambda})^3g_2\right].\label{A_eq}
\end{eqnarray}
In what follow, we will choose the upper signs in order to identify
the trivial critical point with the UV fixed point of the RG flow.
\\
\indent As in other cases, $W=|\mc{W}|$ provides the ``real
superpotential'' in term of which the scalar potential can be
written as
\begin{equation}
V=-\frac{1}{6}\left(\frac{\pd W}{\pd
\lambda}\right)^2-\frac{3}{2}W^2\, .
\end{equation}
In the present case, the scalar kinetic terms are given by
\begin{equation}
-\frac{1}{2}P_\mu^{iA}P^{\mu}_{Ai}=-\frac{3}{2}e^{-4\lambda}(e^{4\lambda}-1)^2\varphi'^2-\frac{3}{2}\lambda'^2\,
.
\end{equation}
With all these results, it can be verified that the second order
field equations are satisfied by the first order BPS equations
\eqref{lambda_eq} and \eqref{A_eq}.
\\
\indent We now solve for the RG flow solution. Equation
\eqref{lambda_eq} clearly admits two fixed points at $\lambda=0$ and
$\lambda=\frac{1}{2}\ln \left[\frac{g_2-g_1}{g_2+g_1}\right]$. These
are supersymmetric critical points identified previously. The
solution for equation \eqref{lambda_eq} is given by
\begin{equation}
g_1g_2r=C_1+2g_1\tan^{-1}e^\lambda-2\sqrt{g_2^2-g_1^2}\tanh^{-1}\left[e^\lambda\sqrt{\frac{g_2+g_1}{g_2-g_1}}\right]
+g_2\ln\left[\frac{1+e^\lambda}{1-e^\lambda}\right]
\end{equation}
where the constant $C_1$ can be set to zero by shifting the $r$
coordinate. By choosing $g_1,g_2>0$, it can be seen that as
$\lambda\rightarrow 0$, we find $r\rightarrow \infty$, and
$r\rightarrow -\infty$ as $\lambda\rightarrow \frac{1}{2}\ln
\left[\frac{g_2-g_1}{g_2+g_1}\right]$. These correspond to the UV
and IR fixed points of the RG flow, respectively. Near the two
critical points, we find
\begin{eqnarray}
\textrm{UV}&:&\qquad \lambda\sim e^{-g_1r}\sim e^{-\frac{r}{L_{\textrm{UV}}}}\nonumber \\
\textrm{IR}&:&\qquad \lambda\sim
e^{\frac{g_1g_2}{\sqrt{g_2^2-g_1^2}}r}\sim
e^{\frac{r}{L_{\textrm{IR}}}}\, .\label{UV_IR_N3_flow}
\end{eqnarray}
Therefore, the flow is driven by an operator of dimension
$\Delta=1,2$, and this operator becomes irrelevant in the IR with
the corresponding scaling dimension $\Delta=4$.
\\
\indent Finally, by combining equations \eqref{lambda_eq} and
\eqref{A_eq}, we obtain
\begin{equation}
\frac{dA}{d\lambda}=-\frac{(1+e^{2\lambda})^3g_1+(e^{2\lambda}-1)^3g_2}{(e^{4\lambda}-1)
\left[(1+e^{2\lambda})g_1+(e^{2\lambda}-1)g_2\right]}
\end{equation}
whose solution is given by
\begin{equation}
A=C_2+\lambda-\ln(1-e^{4\lambda})+\ln\left[g_1(1+e^{2\lambda})+g_2(e^{2\lambda}-1)\right].
\end{equation}
The integration constant $C_2$ can be neglected by rescaling the
coordinates of $dx^2_{1,2}$. It can readily be verified that
$A\rightarrow \frac{r}{L}$ when $\lambda\rightarrow
0,\frac{1}{2}\ln\left[\frac{g_2-g_1}{g_2+g_1}\right]$ as expected
for the two conformal fixed points.
\\
\indent We now identify a possible dual operator driving this flow. From the results of \cite{N3_spectrum2,Ring_N3_superfield}, the eight vector multiplets in the $N=3$ gauged supergravity correspond to the global $SU(3)$ flavor current given, in term of the $N=2$ language, by the superfield
\begin{equation}
\Sigma^i_{\phantom{i}j}=\frac{1}{\sqrt{2}}\textrm{Tr}(U^i\bar{U}_j+\bar{V}^iV_j)-\textrm{flavor trace}\, .
\end{equation}
The trace (Tr) above is over the gauge group $SU(N)\times SU(N)$ under which $U^i$ and $V_i$ transform as a bifundamental. The hypermultiplets $(U^i,i\bar{V}^i)$ form a doublet of $SU(2)_R$ and transform in a fundamental representation the $SU(3)$ flavor. The flow given above is driven by scalar fields in the vector multiplets, and these scalars arise from the eleven-dimensional metric rather than the three-form field \cite{N3_spectrum1}. According to the UV behavior in \eqref{UV_IR_N3_flow}, we then expect that the flow is driven by turning on an $SO(3)\times U(1)$ invariant combination of the scalar mass terms within $\Sigma^i_{\phantom{i}j}$.    

\section{Flows to non-conformal field theories}\label{Non-conformal}
In this section, we consider RG flows to non-conformal field
theories. The supergravity solutions will interpolate between UV
$AdS_4$ critical points and domain walls in the IR.

\subsection{Flows within $SO(2)\times SO(2)\times SO(2)$ singlet scalars}
We first consider scalars invariant under $SO(2)\times SO(2)\times
SO(2)$ symmetry. The first $SO(2)$ is embedded in the $SO(3)_R$ such
that $\mathbf{3}\rightarrow \mathbf{2}+\mathbf{1}$. From the
branching of $\mathbf{3}_0+\mathbf{1}_0$ in \eqref{SU3_8_decom}
under $SO(2)\subset SO(3)\subset SU(3)$, we find
$\mathbf{2}_0+\mathbf{1}_0+\mathbf{1}_0$. Combining the two
decompositions together, we finally obtain the relevant scalar
representations under $SO(2)\times SO(2)\times SO(2)$
\begin{equation}
2\times [(\mathbf{3}_0,\mathbf{3}_0+\mathbf{1}_0)]=2\times
[2(\mathbf{1},\mathbf{1})_0,(\mathbf{1},\mathbf{2})_0,
2(\mathbf{2},\mathbf{1})_0, (\mathbf{2},\mathbf{2})_0].
\end{equation}
There are accordingly four singlets corresponding to the non-compact
generators
\begin{eqnarray}
\tilde{Y}_1&=&e_{3,11}+e_{11,3},\qquad \tilde{Y}_2=ie_{11,3}-ie_{3,11},\nonumber \\
\tilde{Y}_3&=&e_{3,6}+e_{6,3},\qquad \tilde{Y}_4=ie_{6,3}-ie_{3,6}\,
.
\end{eqnarray}
It should be noted that $\tilde{Y}_{1,2}$ are invariant under a
bigger symmetry $SO(2)\times SU(2)\times U(1)$. The above four
singlets correspond to non-compact directions of $SU(2,1)\subset
SU(3,8)$. We then effectively need to parametrize the
$SU(2,1)/SU(2)\times U(1)$ coset manifold. It is more convenient to
adopt a parametrization using $SU(2)$ Euler angles. The $SU(2)\times
U(1)$ compact subgroup of the $SU(2,1)$ group is generated by
\begin{eqnarray}
J_1&=&\frac{i}{2}(e_{11,11}-e_{66}),\qquad J_2=\frac{1}{2}(e_{6,11}-e_{11,6}), \nonumber \\
J_3&=&-\frac{i}{2}(e_{6,11}+e_{11,6}),\qquad
\hat{J}=\frac{i}{2\sqrt{3}}(2e_{33}-e_{66}-e_{11,11})
\end{eqnarray}
with
$\left[J_\alpha,J_\beta\right]=\epsilon_{\alpha\beta\gamma}J_\gamma$
and $\hat{J}$ corresponding to the $U(1)$.
\\
\indent The coset representative for $SO(2)\times SO(2)\times SO(2)$
invariant scalars is accordingly parametrized by
\begin{equation}
L=e^{\varphi_1 J_1}e^{\varphi_2 J_2}e^{\varphi_3 J_3}e^{\Phi
\tilde{Y}_1}e^{-\varphi_3 J_3}e^{-\varphi_2 J_2}e^{-\varphi_1 J_1}\,
.
\end{equation}
The scalar potential turns out to be independent of all the
$\varphi_i$
\begin{equation}
V=-\frac{1}{2}g_1^2[1+2\cosh(2\Phi)]\label{SO2_3_potential}
\end{equation}
which clearly has only the trivial critical point at $\Phi=0$.
\\
\indent The matrix $S_{AB}$ in this case is diagonal
\begin{equation}
S_{AB}=\frac{1}{2}g_1\cosh\Phi\delta_{AB}
\end{equation}
implying that the maximal $N=3$ supersymmetry is preserved if the
conditions $\delta\lambda_i=0$ and $\delta\lambda_{iA}=0$ can be
satisfied. This is similar to solutions studied in the maximal $N=8$
gauged supergravity in \cite{Pope_warner_Dielectric_flow}.
\\
\indent We can proceed as in the previous section to analyze other
BPS equations. Since $\mc{W}$ is real in this case, we simply have
$\omega=0$ and $e^{i\Lambda}=\pm 1$. Generally, the flow equations
for a scalar $\phi_i$ is, up to a numerical factor, given by
$G^{ij}\frac{\pd W}{\pd \phi^j}$ in which $G^{ij}$ being the inverse
of the scalar matrix appearing in the scalar kinetic terms. The
above superpotential depending only on $\Phi$ will immediately give
$\varphi_i'=0$. Remarkably, this precisely agrees with the results
from solving $\delta\lambda_i=0$ and $\delta\lambda_{iA}=0$
equations. This is another consistency check for our results.
\\
\indent We now give the flow equations after choosing a choice of
signs such that the $SO(3)\times SU(3)$ $AdS_4$ critical point is
identified with $r\rightarrow \infty$
\begin{eqnarray}
\Phi'&=&-g_1\sinh \Phi,\qquad \varphi_i'=0,\qquad i=1,2,3,\nonumber \\
A'&=&g_1\cosh \Phi\, .
\end{eqnarray}
A solution to the above equations can be readily obtained
\begin{equation}
\Phi=\pm\ln\left[\frac{e^{g_1r}-e^{C}}{e^{g_1r}+e^C}\right],\qquad
A=\Phi-\ln(1-e^{2\Phi})+C'\, .
\end{equation}
\indent As $r\rightarrow\infty$, the solution approaches the UV
$AdS_4$ critical point with $\Phi\sim e^{-g_1r}$ and $A\sim g_1r$.
At $g_1r\sim C$, there is a singularity with $\Phi$ becoming
infinite
\begin{equation}
\Phi\sim \pm\ln(g_1r-C).
\end{equation}
Both of the signs give rise to the same domain wall metric in the IR
\begin{equation}
ds^2=(g_1r-C)^2dx^2_{1,2}+dr^2\, .
\end{equation}
It can also be checked that the potential \eqref{SO2_3_potential} is
bounded above for $\Phi\rightarrow \pm \infty$ namely
$V(\Phi\rightarrow\pm \infty)\rightarrow -\infty$. The singularity
is then physical according to the criterion of
\cite{Gubser_singularity}. Therefore, the solution should be
interpreted as an RG flow from the UV $N=3$ SCFT to an $N=3$
non-conformal field theory in the IR.

\subsection{Flows within $SO(2)_{\textrm{diag}}\times SO(2)$ singlet scalars}
The solutions considered in the previous subsection describe RG
flows from the trivial $N=3$ critical point. These solutions do not
connect to the non-trivial $AdS_4$ critical point identified in
section \ref{AdS4_flow}. We now consider another class of flow
solutions describing RG flows from both the trivial and non-trivial
critical points to IR gauge theories with $SO(2)\times SO(2)$
symmetry.
\\
\indent We will consider scalars which are singlets under
$SO(2)_{\textrm{diag}}\times SO(2)\subset SO(2)\times SO(2)\times
SO(2)$ symmetry. Further decomposing the scalar representations
gives eight singlets under this symmetry. These correspond to the
following $SU(3,8)$ non-compact generators
\begin{eqnarray}
\bar{Y}_1&=&e_{36}+e_{63},\qquad
\bar{Y}_2=-ie_{36}+ie_{63},\nonumber \\
\bar{Y}_3&=&e_{25}+e_{52}+e_{14}+e_{41},\qquad
\bar{Y}_4=-ie_{25}+ie_{52}-ie_{14}+ie_{41},\nonumber \\
\bar{Y}_5&=&e_{15}+e_{51}-e_{24}-e_{42},\qquad
\bar{Y}_6=-ie_{15}+ie_{51}+ie_{24}-ie_{42},\nonumber \\
\bar{Y}_7&=&e_{3,11}+e_{11,3},\qquad
\bar{Y}_8=-ie_{3,11}+ie_{11,3}\, .
\end{eqnarray}
\indent In this case, using Euler parametrization does not simplify
the result to any useful extent. We then simply parametrize the
coset representative in a straightforward way
\begin{equation}
L=e^{\Phi_1\bar{Y}_1}e^{\Phi_2\bar{Y}_2}e^{\Phi_3\bar{Y}_3}
e^{\Phi_4\bar{Y}_4}e^{\Phi_5\bar{Y}_5}e^{\Phi_6\bar{Y}_6}e^{\Phi_7\bar{Y}_7}
e^{\Phi_8\bar{Y}_8}\, .\label{L_SO2D}
\end{equation}
The resulting scalar potential and BPS equations are much more
complicated than the previous cases. We refrain from giving their
explicit form here.
\\
\indent However, there are some interesting truncations. We will
simply consider these and give the full result within these
truncations. With only $\Phi_7$ and $\Phi_8$ non-vanishing, the
residual symmetry is enhanced to $SO(2)\times SU(2)\times SO(2)$.
Furthermore, if one of these two scalars is set to zero, we recover
the result obtained in the previous subsection. A new deformation
arises from $\Phi_7$ and $\Phi_8$ both being non-zero. In this case,
the $N=3$ supersymmetry is broken to $N=1$.
\\
\indent The matrix $S_{AB}$ is diagonal with two different
eigenvalues, with $S_{11}=S_{22}$. It turns out that the third
eigenvalue gives the true superpotential
\begin{equation}
\mc{W}=2S_{33}=g_1\cosh\Phi_7\cosh\Phi_8+ig_1\sinh\Phi_7\sinh\Phi_8
\end{equation}
in tern of which the scalar potential can be written as
\begin{eqnarray}
V&=&\frac{1}{2}G^{ij}\frac{\pd W}{\pd \Phi^i}\frac{\pd
W}{\pd \Phi^j}-\frac{3}{2}W^2\nonumber \\
&=&-\frac{1}{2}g_1^2[1+2\cosh(2\Phi_7)\cosh(2\Phi_8)]
\end{eqnarray}
where the real superpotential is given by
\begin{equation}
W=|\mc{W}|=\frac{1}{\sqrt{2}}g_1\sqrt{1+\cosh(2\Phi_7)\cosh(2\Phi_8)}\,
.
\end{equation}
In the above result, we have used the scalar kinetic term
\begin{equation}
-\frac{1}{2}G_{ij}\pd_\mu \Phi^i\pd^\mu
\Phi^j=-\frac{1}{2}P_\mu^{Ai}P^\mu_{iA}=-\frac{1}{2}\cosh^2(2\Phi_8)\Phi_7'^2-\frac{1}{2}\Phi_8'^2
\end{equation}
which gives $G_{ij}$, $i, j=7,8$. The inverse $G^{ij}$ can readily
be read off. The supersymmetry transformations of $\psi_{\mu A}$
corresponding to $\epsilon_{1,2}$ can be satisfied by setting
$\epsilon_{1,2}=0$. Accompanied by the usual $\gamma^r$ projection,
the unbroken supersymmetry is then $N=1$ Poincare supersymmetry in
three dimensions.
\\
\indent The BPS equations coming from $\delta\lambda_{iA}=0$ has no
components along $\epsilon_3$. They are accordingly automatically
satisfied with $\epsilon_{1,2}=0$. $\delta\lambda_i=0$ equations
become
\begin{equation}
e^{i\Lambda}[\cosh(2\Phi_8)\Phi_7'+i\Phi_8']=g_1\cosh\Phi_8\sinh\Phi_7+ig_1\cosh\Phi_7\sinh\Phi_8\,
.
\end{equation}
By a similar analysis as in the previous section, we find
$e^{i\Lambda}=\pm e^{i\omega}$ with
$e^{i\omega}=\frac{\mc{W}}{|\mc{W}|}$. The above equations can be
solved by
\begin{eqnarray}
\Phi_7'&=&\mp
\frac{g_1\sinh(2\Phi_7)\textrm{sech}(2\Phi_8)}{\sqrt{2+2\cosh(2\Phi_7)\cosh(2\Phi_8)}},\\
\Phi_8'&=&\mp
\frac{g_1\sinh(2\Phi_8)\cosh(2\Phi_7)}{\sqrt{2+2\cosh(2\Phi_7)\cosh(2\Phi_8)}}\, .
\end{eqnarray}
Together with $A'=\pm W$, these form the full set of flow equations.
\\
\indent By combining these equations, we can solve for $\Phi_8$ and
$A$ as a function of $\Phi_7$
\begin{eqnarray}
\coth(2\Phi_8)&=&\textrm{csch}(2\Phi_7),\\
A&=&-\frac{1}{2}\tanh^{-1}\left[\frac{\sqrt{2}\cosh(2\Phi_7)}{\sqrt{3-\cosh(4\Phi_7)}}\right]
-\frac{1}{4}\ln [\cosh(4\Phi_7)-3]\nonumber
\\
& &+\frac{1}{2}\ln\sinh(2\Phi_7).
\end{eqnarray}
In principle, we can put the solution for $\Phi_8$ in $\Phi_7'$
equation and solve for $\Phi_7(r)$. However, we have not found the
full analytic solution for $\Phi_7(r)$. In the following, we simply
study the $\Phi_7$ behaviors near the UV $AdS_4$ critical point and
near the IR singularity. As $r\rightarrow \infty$, we find
\begin{equation}
\Phi_7\sim \Phi_8\sim e^{-g_1r},\qquad A\sim g_1r\, .
\end{equation}
At large $|\Phi_7|$, we find
\begin{eqnarray}
& &\Phi_7\sim \pm\frac{1}{3}\ln (g_1r),\qquad \Phi_8\sim
\textrm{constant},\nonumber \\
& &ds^2=(g_1r)^{\frac{2}{3}}dx^2_{1,2}+dr^2
\end{eqnarray}
where we have put the singularity at $r=0$ by choosing an
integration constant. These singularities are also physical. The
solution preserves two supercharges and describes an RG flow from
$N=3$ SCFT to a non-conformal field theory in the IR with $N=1$
supersymmetry.
\\
\indent We will now make another truncation by setting $\Phi_i=0$
for $i=2,4,6,8$. This can be verified to be consistent with both the
BPS equations and the second order field equations. In this
truncation, the scalar potential is given by
\begin{eqnarray}
V&=&-\frac{1}{64}\left[(1+\cosh(2\Phi_3)\cosh(2\Phi_5))\left[4\cosh(2\Phi_1)-4+3\cosh[2(\Phi_1-\Phi_3-\Phi_5)]
\right.
\right.\nonumber \\
& &+2\cosh[2(\Phi_1+\Phi_3-\Phi_5)]+3\cosh[2(\Phi_1-\Phi_3+\Phi_5)]
+2\cosh[2(\Phi_3+\Phi_5)] \nonumber \\
&
&\left.+3\cosh[2(\Phi_1+\Phi_3+\Phi_5)]+8\cosh^2\Phi_1[1+3\cosh(2\Phi_3)
\cosh(2\Phi_5)]\cosh(2\Phi_7)\right]g_1^2\nonumber \\
& &-12(\cosh(4\Phi_3)+2\cosh^2(2\Phi_3)\cosh(4\Phi_5)-3)\cosh^2\Phi_7\sinh(2\Phi_1)g_1g_2\nonumber \\
&
&+[\cosh(2\Phi_3)\cosh(2\Phi_5)-1]\left[4+4\cosh(2\Phi_1)-3\cosh[2(\Phi_1-\Phi_3-\Phi_5)]\right.\nonumber
\\
& &
+2\cosh[2(\Phi_3-\Phi_5)]-3\cosh[2(\Phi_1+\Phi_3-\Phi_5)]-3\cosh[2(\Phi_1-\Phi_3+\Phi_5)]\nonumber \\
&
&+2\cosh[2(\Phi_3+\Phi_5)]-3\cosh[2(\Phi_1+\Phi_3+\Phi_5)]\nonumber \\
& &\left.\left.+8[1-3\cosh(2\Phi_3)\cosh(2\Phi_5)]
\cosh(2\Phi_7)\sinh^2\Phi_1\right]g_2^2\right].\label{V_SO2_3}
\end{eqnarray}
Using the same procedure as before, we find the full set of the BPS
equations within this particular truncation
\begin{eqnarray}
\Phi_1'&=&-\frac{1}{8}\frac{e^{-\Phi_1-2(\Phi_3+\Phi_5+\Phi_7)}}{1+e^{2\Phi_7}}
\left[(e^{2\Phi_1}-1)(1+e^{4\Phi_3}+e^{4\Phi_5}+4e^{2(\Phi_3+\Phi_5)}+e^{4(\Phi_3+\Phi_5)})g_1\right.\nonumber
\\
&
&\left.+(1+e^{2\Phi_1})(1+e^{4\Phi_3}+e^{4\Phi_5}-4e^{2(\Phi_3+\Phi_5)}+e^{4(\Phi_3+\Phi_5)})g_2\right],\\
\Phi_3'&=&-\frac{e^{-\Phi_1-2\Phi_3+2\Phi_5-\Phi_7}}{8(1+e^{4\Phi_5})}(e^{4\Phi_3}-1)(1+e^{2\Phi_7})
\left[(1+e^{2\Phi_1})g_1+(e^{2\Phi_1}-1)g_2\right],\\
\Phi_5'&=&-\frac{1}{32}e^{-\Phi_1-2\Phi_3-2\Phi_5-\Phi_7}(e^{4\Phi_3}+1)(1+e^{2\Phi_7})(e^{4\Phi_5}-1)\times
\nonumber \\
& &
\left[(1+e^{2\Phi_1})g_1+(e^{2\Phi_1}-1)g_2\right],\\
\Phi_7'&=&-\frac{1}{32}e^{-\Phi_1-2\Phi_3-2\Phi_5-\Phi_7}(1-e^{2\Phi_7})
\left[(e^{2\Phi_1}-1)\left[1+e^{4\Phi_3}+e^{4\Phi_5}+4e^{2(\Phi_3+\Phi_5)}\right.\right.\nonumber
\\
&
&\left.\left.+e^{4(\Phi_3+\Phi_5)}\right]g_1+(1+e^{2\Phi_1})\left[1+e^{4\Phi_3}+e^{4\Phi_5}-4e^{2(\Phi_3+\Phi_5)}
 +e^{4(\Phi_3+\Phi_5)}\right]g_2\right],\nonumber \\
 & &\\
A'&=&\frac{1}{32}e^{-\Phi_1-2\Phi_3-2\Phi_5-\Phi_7}(1+e^{2\Phi_7})
\left[(e^{2\Phi_1}-1)\left[1+e^{4\Phi_3}+e^{4\Phi_5}+4e^{2(\Phi_3+\Phi_5)}\right.\right.\nonumber
\\
& &\left.\left.+e^{4(\Phi_3+\Phi_5)}\right]g_1+(1+e^{2\Phi_1})
\left[1+e^{4\Phi_3}+e^{4\Phi_5}-4e^{2(\Phi_3+\Phi_5)}+e^{4(\Phi_3+\Phi_5)}\right]g_2\right].\nonumber
\\
& &
\end{eqnarray}
Due to the $\gamma_r$ projector, the solutions will preserve six
supercharges or $N=3$ supersymmetry in three dimensions. When
$\Phi_3=\Phi_1$ and $\Phi_5=\Phi_7=0$, the above equations reduce to
those considered in section \ref{AdS4_flow}. These equations do not
admit any non-trivial $AdS_4$ fixed points apart from the $N=3$
$SO(3)_{\textrm{diag}}\times U(1)$ critical point already identified
in section \ref{AdS4_flow}. This agrees with the remark given in
\cite{Castellani_book} in which partial supersymmetry breaking has
been shown to be impossible.
\\
\indent We are now in a position to consider various possible RG
flows from the UV $N=3$ SCFTs. In this case, we have not found any
possible analytic solutions. Therefore, numerical solutions will be
needed in order to obtain the full flow solutions. Although these
solutions always exist and can be found by imposing suitable
boundary conditions, we will not give them here. Instead, we will
give the behavior near the IR singularity which can be put to $r=0$
by choosing appropriate constants of integration. This is similar to
the analysis given in \cite{Warner_N1_nonCFT_flow}. Note also that,
from the above equations, setting $\Phi_5=0$ and $\Phi_7=0$ is also
a consistent truncation.
\\
\indent We will now consider RG flows to the IR with infinite values
of scalar fields. From the above equations, as $\Phi_3\rightarrow
\pm \infty$, we find that $\Phi_5'\rightarrow 0$. Since both of the
$AdS_4$ critical points have $\Phi_5=0$, we will set $\Phi_5=0$
throughout the analysis.
\\
\indent At the trivial $N=3$ $AdS_4$ critical point, all scalars are
dual to relevant operator of dimensions $\Delta=1,2$. For
$\Phi_3>0$, there are flows with the IR behavior
\begin{eqnarray}
& &\Phi_1\sim \phi_0,\qquad \Phi_7\sim \Phi_3,\qquad
\Phi_3\sim-\frac{1}{3}\ln
\left[\frac{3}{8}\tilde{g}r\right],\nonumber \\
& &ds^2=r^{\frac{2}{3}}dx^2_{1,2}+dr^2
\end{eqnarray}
where $\phi_0$ is a constant and
$\tilde{g}=g_1\cosh\phi_0+g_2\sinh\phi_0$. There is also another
flow with asymptotic behavior
\begin{eqnarray}
& &\Phi_1\sim \phi_0,\qquad \Phi_7\sim -2\Phi_3,\qquad
\Phi_3\sim-\frac{1}{4}\ln
\left[\frac{1}{2}\tilde{g}r\right],\nonumber \\
& &ds^2=r^{\frac{1}{2}}dx^2_{1,2}+dr^2\, .
\end{eqnarray}
\indent For $\Phi_3<0$, we have flows with
\begin{eqnarray}
& &\Phi_1\sim \phi_0,\qquad \Phi_7\sim \pm\Phi_3,\qquad
\Phi_3\sim-\frac{1}{3}\ln
\left[\frac{3}{8}\tilde{g}r\right],\nonumber \\
& &ds^2=r^{\frac{2}{3}}dx^2_{1,2}+dr^2\, .
\end{eqnarray}
It should be noted that when $\Phi_7\neq 0$, we always have constant
$\Phi_1$ in the IR. This is however not the case when $\Phi_7=0$. An
example of this flow is given by
\begin{eqnarray}
& &\Phi_1\sim -2\Phi_3,\qquad \Phi_7=0,\qquad
\Phi_3\sim-\frac{1}{4}\ln
\left[\frac{1}{2}(g_1-g_2)r\right],\nonumber \\
& &ds^2=r^{\frac{1}{2}}dx^2_{1,2}+dr^2\, .
\end{eqnarray}
Remarkably, all of these flows are physical according to the
criterion of \cite{Gubser_singularity} as can be checked from
\eqref{V_SO2_3} that all the flows give $V\rightarrow -\infty$.
\\
\indent The non-trivial $AdS_4$ critical point can be approached by
setting $\Phi_1=\pm
\Phi_3=\Phi_0=\frac{1}{2}\ln\left[\frac{g_2-g_1}{g_2+g_1}\right]$ in
the UV with different signs corresponding to different combinations
of $SO(3)\times SO(3)$ generators in forming
$SO(3)_{\textrm{diag}}$. We will additionally set $\Phi_7=\Phi_5=0$
in the following analysis.
\\
\indent For $\Phi_3>\Phi_0$, there is a flow with asymptotic
behavior
\begin{eqnarray}
& &\Phi_1\sim \Phi_3\sim-\frac{1}{3}\ln
\left[\frac{3}{8}(g_1+g_2)r\right],\nonumber \\
& &ds^2=r^{\frac{2}{3}}dx^2_{1,2}+dr^2\, .
\end{eqnarray}
\indent For $\Phi_3<\Phi_0$, we have flows with the IR behavior
\begin{eqnarray}
& &\Phi_1\sim \pm\Phi_3,\qquad \Phi_3\sim\frac{1}{3}\ln
\left[\frac{3}{8}(g_1\mp g_2)r\right],\nonumber \\
& &ds^2=r^{\frac{2}{3}}dx^2_{1,2}+dr^2\, .
\end{eqnarray}
All of these flows are also physical with $V\rightarrow -\infty$
near the IR singularity.

\section{Flows to lower dimensions}\label{Lower-dimensional flow}
In this section, we consider supersymmetric solutions of the form
$AdS_2\times \Sigma_2$ in which $\Sigma_2$ is a Riemann surface in
the form of a two-sphere $S^2$ or a two-dimensional hyperbolic space
$H^2$. Domain wall solutions interpolating between $AdS_4$ critical
points and these geometries should be interpreted as RG flows to
lower dimensional superconformal field theories. In the present
case, the lower dimensional SCFTs would be described by twisted
compactifications of the $N=3$ SCFTs in three dimensions resulting
in one-dimensional SCFTs. We will look for supersymmetric $AdS_2$
solutions with $SO(2)\times SO(2)\times SO(2)$ and $SO(2)\times
SO(2)$ symmetries within $N=3$ $SO(3)\times SU(3)$ gauged
supergravity.

\subsection{$AdS_2$ critical points with $SO(2)\times SO(2)\times SO(2)$ symmetry}
We begin with the BPS equations relevant for the present analysis.
The gauge fields are now non-vanishing. We adopt the twist procedure
in order to preserve some amount of supersymmetry. This involves
turning on some gauge field to cancel the spin connection along the
$\Sigma_2$ directions. We will primarily consider the case of curved
$\Sigma_2$ in the form of $S^2$ and $H^2$.
\\
\indent The four-dimensional metric is taken to be
\begin{equation}
ds_4^2=-e^{2A(r)}dt^2+dr^2+e^{2B(r)}ds^2(\Sigma_2)
\end{equation}
where $ds^2(\Sigma_2)$ is the metric on $\Sigma_2$. Its explicit
form can be written as
\begin{equation}
ds^2(S^2)=d\theta^2+\sin^2\theta d\phi^2\qquad \textrm{and}\qquad
ds^2(H^2)=\frac{1}{y^2}(dx^2+dy^2)
\end{equation}
for the $S^2$ and $H^2$ cases, respectively. In the following, we
will only give the detail of the $S^2$ case. The $H^2$ case can be
done in a similar way.
\\
\indent The component of the spin connection on $S^2$ that needs to
be canceled is given by
\begin{equation}
\omega^{\hat{\phi}\hat{\theta}}=e^{-B}\cot\theta e^{\hat{\phi}}.
\end{equation}
This appears in the $\delta \psi_{\phi A}$ variation. To cancel this
contribution, we turn on some of the gauge fields $A_{\mu A}$
appearing in the $SU(3)$ composite connection $Q_A^{\phantom{A}B}$.
We will choose the non-vanishing gauge field to be
\begin{equation}
A_3=a\cos\theta d\phi
\end{equation}
which gives rise to the non-vanishing components of the composite
connection
\begin{equation}
Q_1^{\phantom{A}2}=-Q_2^{\phantom{A}1}=-a_1g_1\qquad
\textrm{or}\qquad Q_{AB}=-g_1\epsilon_{ABC}A^C\, .
\end{equation}
The cancelation is achieved by imposing the following twist and
projection conditions
\begin{equation}
a_1g_1=\frac{1}{2},\qquad
\gamma_{\hat{\phi}\hat{\theta}}\epsilon_a=i\sigma_{2a}^{\phantom{2a}b}\epsilon_b,\qquad
a,b=1,2\, .\label{Twist_condition}
\end{equation}
In the above equation, $\sigma_{2a}^{\phantom{2a}b}$ denotes the
usual second Pauli matrix. We have split the index $A$ into $(a,3)$
such that $\epsilon^A=(\epsilon^a,\epsilon^3)$. It should be noted
that with only $A_3$ non-vanishing, the supersymmetry corresponding
to $\epsilon^3$ cannot be preserved, so we will set $\epsilon^3=0$.
Eventually, there are only four unbroken supercharges corresponding
to $\epsilon^a$ that are subject to the
$\gamma_{\hat{\phi}\hat{\theta}}$ projection.
\\
\indent In addition, there are other two gauge fields that can be
turned on along with $A_3$. These correspond to the $SO(2)\times
SO(2)\subset SU(3)$ symmetry and are given by
\begin{equation}
A^6=b\cos\theta d\phi\qquad \textrm{and}\qquad A^{11}=c\cos\theta
d\phi\, .
\end{equation}
All other gauge fields are zero. The field strengths of
$(A^3,A^6,A^{11})$ are given by
\begin{equation}
F_\Lambda =-a_\Lambda e^{-2B}e^{\hat{\theta}}\wedge e^{\hat{\phi}}
\end{equation}
with non-vanishing $a_\Lambda=(a_3,a_6,a_{11})=(a,b,c)$. With the
convention $\epsilon^{\hat{t}\hat{r}\hat{\theta}\hat{\phi}}=1$, we
find the dual field strength
\begin{equation}
\tilde{F}_\Lambda=a_\Lambda e^{-2B}e^{\hat{t}}\wedge e^{\hat{r}}\, .
\end{equation}
The four-dimensional chirality on $\epsilon_A$ relates the
$\gamma_{\hat{\phi}\hat{\theta}}$ to the $\gamma_{\hat{t}\hat{r}}$
as follow
\begin{equation}
\gamma_5\epsilon_a=i\gamma_{\hat{t}}\gamma_{\hat{r}}\gamma_{\hat{\theta}}\gamma_{\hat{\phi}}\epsilon_a=\epsilon_a
\end{equation}
implying that
\begin{equation}
\gamma_{\hat{t}\hat{r}}\epsilon_{a}=\sigma_{2a}^{\phantom{2a}b}\epsilon_b\,
.
\end{equation}
\indent We now in a position to set up the BPS equations by using
all of the above conditions and the formulae given in section
\ref{N3theory}. In the presence of gauge fields, unlike the
solutions considered in section \ref{AdS4_flow}, it turns out that
the parametrization of the coset representative for $SO(2)\times
SO(2)\times SO(2)$ invariant scalars using $SU(2)$ Euler angles does
not simplify the resulting equations to any appreciable degree. We
will rather choose to parametrize the coset representative in the
form of
\begin{equation}
L=e^{\tilde{Y}_1\Phi_1}e^{\tilde{Y}_2\Phi_2}e^{\tilde{Y}_3\Phi_3}e^{\tilde{Y}_4\Phi_4}\,
.
\end{equation}
Furthermore, we will make a truncation $\Phi_2=\Phi_4=0$ to make
things more manageable. This can also be verified to be consistent
with all of the BPS equations as well as the corresponding field
equations.
\\
\indent As in the previous cases, the equations coming from
$\delta\chi=0$ are identically satisfied since
$C_{M}^{\phantom{M}MA}=0$, and the particular ansatz for the gauge
fields given above gives $G^A_{\mu\nu}\gamma^{\mu\nu}=0$. In
addition, $\delta\lambda_i=0$ equations are identically satisfied
provided that we set $\epsilon^3=0$. In our particular truncation,
$\mc{W}$ is real, so we can impose the $\gamma_{\hat{r}}$ projection
simply as $\gamma_{\hat{r}}\epsilon_a=\pm \epsilon_a$. With the
usual choice of signs chosen, the independent BPS equations coming
from $\delta \lambda_{Ai}=0$ are given by
\begin{eqnarray}
\Phi_1'&=&\frac{1}{4}e^{-\Phi_1-\Phi_3-2B}\left[4ce^{\Phi_3}(1+e^{2\Phi_1})+2b(e^{2\Phi_1}-1)(e^{2\Phi_3}-1)
\right.\nonumber \\
& &\left.-2a(e^{2\Phi_1}-1)(1+e^{2\Phi_3})+g_1e^{2B}(1-e^{2\Phi_1})+g_1e^{2\Phi_3+2B}(1-e^{2\Phi_1})\right],\,\,\,\\
\Phi_3'&=&-\frac{e^{\Phi_1-\Phi_3-2B}}{1+e^{2\Phi_1}}\left[2a(e^{2\Phi_3}-1)-2b(1+e^{2\Phi_3})+g_1e^{2B}
(e^{2\Phi_3}-1)\right].
\end{eqnarray}
\indent With the twist conditions \eqref{Twist_condition},
$\delta\psi_{\hat{\phi}A}=0$ equations are the same as
$\delta\psi_{\hat{\theta}A}=0$ equations. All of these conditions
reduce to a single equation for the function $B$ while the
conditions $\delta\psi_{\mu A}$, for $\mu=t$, give an equation for
the function $A$. These are given by
\begin{eqnarray}
B'&=&-\frac{1}{4}e^{-\Phi_1-\Phi_3-2B}\left[2ce^{\Phi_3}(1-e^{2\Phi_1})-b(1+e^{2\Phi_1})(e^{2\Phi_3}-1)
\right. \nonumber \\
& &\left.+a(1+e^{2\Phi_1})(1+e^{2\Phi_3})-g_1e^{2B}(1+e^{2\Phi_1})-g_1e^{2\Phi_3+2B}(1+e^{2\Phi_1})\right],\\
A'&=&-\frac{1}{4}e^{-\Phi_1-\Phi_3-2B}\left[-2ce^{\Phi_3}(1-e^{2\Phi_1})+b(1+e^{2\Phi_1})(e^{2\Phi_3}-1)
\right. \nonumber \\
&
&\left.-a(1+e^{2\Phi_1})(1+e^{2\Phi_3})-g_1e^{2B}(1+e^{2\Phi_1})-g_1e^{2\Phi_3+2B}(1+e^{2\Phi_1})\right].\,\,\,\,\,\,
\end{eqnarray}
For the $H^2$ case, a similar analysis can be carried out. The
result is the same as the above equations with $(a,b,c)$ replaced by
$(-a,-b,-c)$.
\\
\indent An $AdS_2\times \Sigma_2$ geometry is given by a fixed point
of the above equations satisfying $\Phi_1'=\Phi_3'=B'=0$ and
$A'=\frac{1}{L_{AdS_2}}$. We find a class of solutions given by
\begin{eqnarray}
B&=&\frac{1}{2}\ln\left[\frac{2[a(1-e^{2\Phi_3})+b(1+e^{2\Phi_3})]}{(e^{2\Phi_3}-1)g_1}\right],\nonumber \\
\Phi_1&=&\frac{1}{2}\ln\left[\frac{c(1-e^{2\Phi_3})-2be^{\Phi_3}}{c(e^{2\Phi_3}-1)-2be^{\Phi_3}}\right],\nonumber \\
\Phi_3&=&\frac{1}{2}\ln \left[\frac{b^2-c^2\pm
\sqrt{b^2[9a^2-8(b^2+c^2)]}}{3ab-3b^2-c^2}\right].\label{AdS2_SO2_3}
\end{eqnarray}
The expression for the $AdS_2$ radius is much more complicated. We
will not give it here, but in any case this can be obtained by
substituting the values of $B$, $\Phi_1$ and $\Phi_3$ in the $A'$
equation.

\subsection{$AdS_2$ critical points with $SO(2)\times SO(2)$ symmetry}
We now look for $AdS_2$ solutions that can be obtained from twisted
compactifications of the non-trivial $AdS_4$ critical point. As in
section \ref{Non-conformal}, we consider
$SO(2)_{\textrm{diag}}\times SO(2)$ invariant scalars. The coset
representative is still given by \eqref{L_SO2D}. The ansatze for the
gauge fields are similar to the previous case but with
$b=\frac{g_1}{g_2}a$ to implement the gauge field of
$SO(2)_{\textrm{diag}}$.
\\
\indent Following the same procedure as in the previous subsection,
we obtain a set of BPS equations, again in a consistent truncation
with $\Phi_i=0$, for $i=2,4,6,8$,
\begin{eqnarray}
\Phi_1'&=&-\frac{2e^{\Phi_7}}{1+e^{2\Phi_7}}\left[-\frac{a}{g_2}e^{-\Phi_1-2B}[(1+e^{2\Phi_1})g_1
+(1-e^{2\Phi_1})g_2]\right.\nonumber \\
&
&+\frac{1}{16}e^{-\Phi_1-2\Phi_3-2\Phi_5}\left[(e^{2\Phi_1}-1)(1+e^{4\Phi_3}+e^{4\Phi_5}
+4e^{2(\Phi_3+\Phi_5)}+e^{4(\Phi_3+\Phi_5)})g_1\nonumber\right.
\\
& &\left.\left.+(1+e^{2\Phi_1})(1+e^{4\Phi_3}+e^{4\Phi_5}
-4e^{2(\Phi_3+\Phi_5)}+e^{4(\Phi_3+\Phi_5)})g_2\right]\right],\\
\Phi_3'&=&-\frac{1}{8}\left[\frac{1+e^{2\Phi_7}}{1+e^{4\Phi_5}}\right]
e^{-\Phi_1-2\Phi_3+2\Phi_5-\Phi_7}(e^{4\Phi_3}-1)\times\nonumber \\
& &[(1+e^{2\Phi_1})g_1+(e^{2\Phi_1}-1)g_2],\\
\Phi_5'&=&-\frac{1}{32}e^{-\Phi_1-2\Phi_3-2\Phi_5-\Phi_7}(1+e^{4\Phi_3})(e^{4\Phi_5}-1)(1+e^{2\Phi_7})
\times\nonumber \\
& &[(1+e^{2\Phi_1})g_1+(e^{2\Phi_1}-1)g_2],\\
\Phi_7'&=&\frac{1}{32}e^{-\Phi_1-2\Phi_3-2\Phi_5-\Phi_7}(1-e^{2\Phi_7})
\left[(1+e^{2\Phi_1})\left[1+e^{4\Phi_3}+e^{4\Phi_5}+4e^{2(\Phi_3+\Phi_5)}\right.\right.\nonumber \\
& &\left.\left.+e^{4(\Phi_3+\Phi_5)}\right]g_1
+(e^{2\Phi_1}-1)(1+e^{4\Phi_3}+e^{4\Phi_5}-4e^{2(\Phi_3+\Phi_5)}+e^{4(\Phi_3+\Phi_5)})g_2\right]\nonumber
\\
&
&+\frac{1}{2g_2}e^{-\Phi_1-\Phi_7-2B}\left[2ce^{\Phi_1}(1+e^{2\Phi_7})g_2+a(e^{2\Phi_7}-1)\times
\right.\nonumber \\
& &\left.
[(e^{2\Phi_1}-1)g_1-(1+e^{2\Phi_1})g_2]\right],\\
B'&=&-\frac{1}{32}e^{-\Phi_1-\Phi_7-2B}\left[\frac{8a}{g_2}(1+e^{2\Phi_7})[(1-e^{2\Phi_1})g_1
+(1+e^{2\Phi_1})g_2]\right.\nonumber
\\
&
&-e^{-2(\Phi_3+\Phi_5)}\left[16ce^{\Phi_1+2\Phi_3+2\Phi_5}(e^{2\Phi_7}-1)\right.\nonumber
\\
&
&+e^{2B}(1+e^{2\Phi_7})\left[(1+e^{2\Phi_1})(1+e^{4\Phi_3}+e^{4\Phi_5}+4e^{2(\Phi_3+\Phi_5)}
+e^{4(\Phi_3+\Phi_5)})g_1\right.\nonumber \\
& &
\left.\left.+(e^{2\Phi_1}-1)(1+e^{4\Phi_3}+e^{4\Phi_5}-4e^{2(\Phi_3+\Phi_5)}
+e^{4(\Phi_3+\Phi_5)})g_2\right]\phantom{\frac{1}{1}} \right],\\
A'&=&\frac{1}{32}e^{-\Phi_1-\Phi_7-2B}\left[\frac{8a}{g_2}(1+e^{2\Phi_7})[(1-e^{2\Phi_1})g_1
+(1+e^{2\Phi_1})g_2]\right.\nonumber
\\
&
&-e^{-2(\Phi_3+\Phi_5)}\left[16ce^{\Phi_1+2\Phi_3+2\Phi_5}(e^{2\Phi_7}-1)\right.\nonumber
\\
&
&-e^{2B}(1+e^{2\Phi_7})\left[(1+e^{2\Phi_1})(1+e^{4\Phi_3}+e^{4\Phi_5}+4e^{2(\Phi_3+\Phi_5)}
+e^{4(\Phi_3+\Phi_5)})g_1\right.\nonumber \\
& &
\left.\left.+(e^{2\Phi_1}-1)(1+e^{4\Phi_3}+e^{4\Phi_5}-4e^{2(\Phi_3+\Phi_5)}
+e^{4(\Phi_3+\Phi_5)})g_2\right]\phantom{\frac{1}{1}} \right].
\end{eqnarray}
From these equations, we find a number of $AdS_2\times \Sigma_2$
solutions given below.
\begin{enumerate}
\item For $\Phi_3=\Phi_5=0$, we find a critical point
\begin{eqnarray}
G&=&\frac{1}{2}\ln\left[\frac{2a[(1+e^{2\Phi_1})g_1+(1-e^{2\Phi_1})g_2]}{(e^{2\Phi_1}-1)g_1g_2}\right],\nonumber
\\
\Phi_7&=&\frac{1}{2}\ln\left[\frac{2ae^{\Phi_1}g_1+c(e^{2\Phi_1}-1)g_2}{2ae^{\Phi_1}g_1-c(e^{2\Phi_1}-1)g_2}
\right],\nonumber \\
\Phi_1&=&\frac{1}{2}\ln\left[\frac{a^2g_1^2-c^2g_2^2\pm\sqrt{a^2g_1^2[a^2(9g_2^2-8g_1^2)-8c^2g_2^2]}}
{3a^2g_1(g_2-g_1)-c^2g_2^2}\right].
\end{eqnarray}
\item For $c=0$, $\Phi_7$ can be consistently set to zero. If we further set $\Phi_3=0$, we find the following
critical point
\begin{eqnarray}
\Phi_1&=&\frac{1}{2}\ln \left[\frac{g_1\pm
\sqrt{9g_2^2-8g_1^2}}{3(g_2-g_1)}\right],\nonumber \\
G&=&\frac{1}{2}\ln
\left[\frac{2a(g_2-g_1)\left[2g_1+3g_2\mp\sqrt{9g_2^2-8g_1^2}\right]}{g_1g_2\left[4g_1-3g_2\pm
\sqrt{9g_2^2-8g_1^2}\right]}\right],\nonumber \\
L_{AdS_2}&=&\frac{2g_1+3g_2\mp
\sqrt{9g_2^2-8g_1^2}}{4g_1^2}\sqrt{\frac{3(g_2-g_1)}{g_1\pm
\sqrt{9g_2^2-8g_1^2}}}\, .
\end{eqnarray}
\item For $c=0$ and $\Phi_7=0$ but $\Phi_3\neq 0$, we find a critical
point
\begin{eqnarray}
\Phi_1&=&\frac{1}{2}\ln\left[\frac{g_2-g_1}{g_2+g_1}\right],\nonumber
\\
G&=&\frac{1}{2}\ln\left[\frac{a}{g_1}+\frac{ag_1}{g_2^2}\right],\nonumber
\\
\Phi_3&=&\frac{1}{2}\ln\left[\frac{g_1^4+10g_1^2g_2^2+g_2^4-2\sqrt{5g_1^6g_2^2+26g_1^4g_2^4+5g_1^2g_2^6}}
{g_2^4-g_1^4}\right],\nonumber \\
L_{AdS_2}&=&\frac{\sqrt{g_2^2-g_1^2}}{2g_1g_2}\, .
\end{eqnarray}
\end{enumerate}
It can be checked that all of the above solutions are valid by
choosing suitable choices of the two coupling $(g_1, g_2)$ and the
parameters $(a,c)$ in a manner that is consistent with the twist
condition $2g_1a=1$. For example, taking $b=2c$ and $a=5c$ in the
solution \eqref{AdS2_SO2_3} leads to
\begin{equation}
G=\ln\left[0.927441\sqrt{\frac{a}{g_1}}\right],\qquad
\Phi_1=0.146711,\qquad \Phi_3=0.287363\, .
\end{equation}
There might be more critical points, but we have not found any other
real solutions.
\\
\indent We end this section by a remark on $AdS_2\times T^2$
solutions. Since $T^2$ is flat, the twist is not needed. We will set
$A_3=0$ or equivalently $a=0$. From the above two cases, we have not
found any valid $AdS_2\times T^2$ solutions.

\section{Conclusions}\label{conclusions}
In this paper, we have studied $N=3$ gauged supergravity in four
dimensions with $SO(3)\times SU(3)$ gauge group. We have found a new supersymmetric
$AdS_4$ critical point, with $SO(3)\times U(1)$ symmetry and
unbroken $N=3$ supersymmetry, and given the full mass spectrum of
all $48$ scalars at this critical point. An analytic RG flow
interpolating between this new critical point and the trivial UV
fixed point has also been explicitly given. The flow describes a
supersymmetric deformation by a relevant operator of dimension
$\Delta=1,2$. It would be of particular interest to precisely
identify the dual operator that drives the flow in the dual $N=3$
SCFT. This result provides another example of supersymmetric
deformations of $N=3$ Chern-Simons-Matter gauge theories which might
be useful in the holographic study of ABJM-type theories coupled to
matter multiplets.
\\
\indent In addition, we have studied RG flows to non-conformal $N=3$
gauge theories in three dimensions with $SO(2)\times SU(2)\times
U(1)$ and $SO(2)_{\textrm{diag}}\times SO(2)$ symmetries. In the
former class of solutions, we have found $N=3$ supersymmetric
deformations in the absence of the ``pseudoscalars'' corresponding
to the imaginary part of the complex scalars. When a pseudoscalar is
turned on, the corresponding deformation breaks supersymmetry to
$N=1$. The latter class includes supersymmetric deformations that
break conformal symmetry of the $SO(3)\times U(1)$ $N=3$ SCFT dual
to the non-trivial $AdS_4$ critical point. Remarkably, all of these
solutions have physically acceptable IR singularities. This is due
to the particular form of the scalar potential which is always
bounded above in the scalar sectors considered in this paper. This
is very similar to the solution studied in
\cite{Pope_warner_Dielectric_flow}. These results would be of
particular interest in describing world volume theory of M2-branes
and hopefully in condensed matter physics systems along the line of
\cite{N3_and_QHE}.
\\
\indent The last result of this paper consists of supersymmetric
$AdS_2\times \Sigma_2$ solutions preserving four supercharges or
$N=2$ Poincare supersymmetry in three dimensions. We have given
$AdS_2$ solutions with $SO(2)\times SO(2)\times SO(2)$ and
$SO(2)_{\textrm{diag}}\times SO(2)$ symmetries. In the context of twisted field theories,
these solutions describe possible twisted compactifications of $N=3$
SCFTs dual to the two $AdS_4$ critical points mentioned above. These
should be useful in the context of AdS$_2$/CFT$_1$ correspondence
and black hole physics. It should also be noted that there is no
$AdS_2\times T^2$ solutions within the scalar submanifolds
considered here.
\\
\indent There are many possible future directions to investigate.
Firstly, it is interesting to find whether the new $SO(3)\times U(1)$ critical point and the corresponding RG flows can be uplifted to eleven dimensions. This would give a geometric interpretation to the solutions
obtained here in the context of M-theory in much the same way as the
recent work for the $N=8$ gauged supergravity in \cite{Warner_N8_uplift}. The complete truncation of eleven-dimensional supergravity on $N^{010}$ keeping only $SU(3)$ singlet fields is given in \cite{N010_truncation_Cassani}. However, the result of \cite{N010_truncation_Cassani} obviously cannot be used to uplift the $AdS_4$ critical point and
the RG flows given in this paper since the scalars that transform
non-trivially under the flavor group $SU(3)$ are also turned on. 
\\
\indent
It should be remarked here about the condition $g_2^2>g_1^2$
related to the existence of the $SO(3)\times U(1)$ critical point.
Within the four-dimensional framework, the two coupling constants are completely
free. The consistency of the gauging does not impose any relation between them. On the other hand, from the
eleven-dimensional point of view, the ratio between $g_1$ and $g_2$ should
be fixed since there is no continuous parameter in $N^{010}$. This might indicate that the $SO(3)\times U(1)$ critical point in eleven dimensions does not exist if the condition $g_2^2>g_1^2$ is not satisfied. Alternatively, this critical point might arise from a more complicated compactification. It would be interesting to investigate these issues in more detail. 
\\
\indent
In finding $AdS_2\times \Sigma_2$ solutions, we have truncated out the
pseudoscalars. It would be interesting to investigate their role in
$AdS_2\times \Sigma_2$ backgrounds as well as in the holographic
AdS$_2$/CFT$_1$ context. In particular, finding black hole solutions interpolating between $N=3$ $AdS_4$ and these $AdS_2\times \Sigma_2$ geometries and comparing the black hole entropy with the result from superconformal indices in the dual $N=3$ SCFT, as in the $AdS_4\times S^7$ case studied in \cite{AdS2_BH3}, would provide an example of this study in a less supersymmetric case. The solutions found here would also be useful in this context. We leave all these issues for future investigations.


\begin{acknowledgments}
This work is supported by Chulalongkorn University
through Ratchadapisek Sompoch Endowment Fund under grant
GF-58-08-23-01 (Sci-Super II). The author is grateful to useful
discussions with C. Nunez and D. Cassani and correspondences
from L. Castellani, C. Ahn and J. P. Gauntlett. He would also like to thank Khem
Upathambhakul for collaborating in a related project. The author is also supported by The Thailand Research Fund (TRF) under grant RSA5980037.
\end{acknowledgments}


\end{document}